\documentclass[journal]{IEEEtran}

\usepackage{graphicx}
\usepackage{amsmath}
\usepackage{amsfonts}
\usepackage{amssymb, color}
\usepackage{cite}
\usepackage{bm} 

\usepackage[ruled,vlined]{algorithm2e}
\usepackage{algorithmic}
\usepackage{booktabs}

\usepackage[protrusion=true,expansion=true]{microtype}

\usepackage[ruled,vlined]{algorithm2e}

\usepackage{caption}
\usepackage{subcaption}

\begin{document} 

\title{ Spatially Robust Near-Field SWIPT Using Pinching Antennas: Rate-Energy Tradeoff Bounds }

\author{Zoran Hadzi-Velkov,~\IEEEmembership{Senior Member,~IEEE}, Marija Poposka,~\IEEEmembership{Graduate~Student~Member,~IEEE,} \hspace{25mm}Slavche Pejoski,~\IEEEmembership{Member,~IEEE}, and Arumugam Nallanathan,~\IEEEmembership{Fellow,~IEEE}

\vspace{-7mm}

\thanks{Z. Hadzi-Velkov, M. Poposka and S. Pejoski are with the Faculty of Electrical Engineering and Information Technologies, Ss. Cyril and Methodius University, Skopje, North Macedonia (e-mail: \{zoranhv, poposkam, slavchep\}@feit.ukim.edu.mk).}  
\thanks{A. Nallanathan is with the School of Electronic Engineering and Computer Science, Queen Mary University of London, UK and also with the Department of Electronic Engineering, Kyung Hee University, Yongin-si, Gyeonggi-do 17104, Korea (e-mail: a.nallanathan@qmul.ac.uk).}

}

\markboth{}{Shell \MakeLowercase{\textit{et al.}}: Bare IEEEtran.cls for Journals}
\maketitle

\begin{abstract} 
Pinching Waveguide Antennas (PWAs) offer significant potential for simultaneous wireless information and power transfer (SWIPT) by enabling precise near-field energy focusing. However, existing optimization frameworks are largely point-based (targeting a single coordinate for maximum gain), and thus highly sensitive to positioning errors and mobility, as near-field signals fluctuate significantly even over small spatial displacements. In this paper, we propose a spatially robust design framework based on discrete antenna selection optimized for service area (SA) coverage. Unlike point-based approaches, our model guarantees quality of service within predefined SAs for both information decoding (ID) and energy harvesting (EH) receivers, thereby improving robustness to user displacements. We formulate the problem as a non-convex binary quadratic program aimed at maximizing harvested energy within the EH SA subject to robust rate constraints in the ID SA. To characterize fundamental performance limits, we develop a semidefinite relaxation (SDR) framework that provides an upper bound on the achievable rate-energy (R-E) region. For the lower bound, we employ a low-complexity swap-based local search algorithm enforcing binary hardware constraints. Numerical results demonstrate that the proposed coverage-oriented design yields a robust R-E tradeoff and maintains stable performance across service regions, highlighting the advantages of discrete antenna activation over point-based near-field optimization approaches. 

\end{abstract}


\vspace{-4mm}

\section{Introduction}

\vspace{-1mm}

Pinching waveguide antennas (PWAs) have recently emerged as a promising enabler for near-field wireless systems, offering a flexible and low-complexity means of shaping electromagnetic radiation through the selective activation of pinching elements along a dielectric waveguide [\ref{docomo_lit}]. By exploiting near-field propagation and spherical wavefronts, PWAs can achieve highly focused energy delivery and information transmission, making them particularly attractive for simultaneous wireless information and power transfer (SWIPT) [\ref{mimo_lit}]. 

Existing research on PWAs has largely focused on optimizing pinching locations to maximize gain at static receiver coordinates [\ref{flexible_lit}]--[\ref{swipt_2_lit}]. While these designs provide high instantaneous performance, they are inherently point-based, targeting users on fixed single-point coordinates, and are thus highly sensitive to positioning uncertainties and user mobility. This vulnerability is especially acute in the near-field, where rapid spatial variations in signal amplitude and phase lead to sharp performance degradation under even slight misalignment [\ref{beamfocus_lit}]. Recent SWIPT extensions have characterized rate-energy (R-E) tradeoffs for static settings, often employing alternating optimization or heuristic searches in multi-user settings [\ref{swipt_1_lit}], [\ref{swipt_2_lit}]. However, these approaches maintain a point-centric focus that is susceptible to near-field spatial variability. To ensure robust performance, optimization should instead target a \textit{service area} (SA) around each receiver, enabling stable operation and guaranteed quality-of-service (QoS) despite user movements.

Motivated by these limitations, this paper proposes a spatially robust SWIPT framework for PWAs. Departing from conventional single-point designs, we optimize antenna activation to guarantee QoS across predefined SAs for both information decoding (ID) and energy harvesting (EH). This region-based formulation inherently mitigates near-field sensitivity and enhances resilience to user mobility. This approach leads to structured binary quadratic programs (BQPs) that enable tight upper bounding based on semidefinite-relaxation (SDR). To evaluate the performance gap associated with binary activation, we employ a swap-based local search algorithm to recover feasible, high-quality integer solutions.

The paper's main contributions are summarized as follows: ($1$) We introduce the spatial covariance matrix for an SA to systematically account for user mobility and spatial uncertainty in pinching beamforming optimization, shifting the design paradigm from single-point gain maximization to guaranteed coverage over a spatial region; ($2$) The resulting beamforming optimizations are structured as BQPs, which enables the novel application of SDR methodology to establish rigorous upper bounds on achievable performance; ($3$) We differentiate between ergodic and pointwise rate QoS requirements, corresponding to average-case and worst-case spatial robustness paradigms, and the resulting R-E tradeoff reveals the performance gap between these two design philosophies. 

\vspace{-1mm}

\section{System Model}

We consider a downlink SWIPT system where a single transmitter, equipped with a pinching waveguide antenna (PWA), simultaneously broadcasts energy and information to an energy harvesting (EH) receiver and an information decoding (ID) receiver, respectively. The receivers are located within a horizontal rectangular service plane, $\mathcal{D} = \{ (x, y, 0) \in \mathbb{R}^3 \, | -D_X \leq x \leq D_X, -D_Y \leq y \leq D_Y \}$, which is discretized into a uniform grid for spatial analysis.

The PWA is deployed along a straight waveguide aligned with the $x$-axis at a fixed height $h$ above the service plane. The waveguide provides $M$ equidistant candidate locations for antenna activation, separated by a distance $\Delta$ chosen to be sufficiently large to neglect mutual electromagnetic coupling. The coordinates of the $m$-th candidate location ($1 \leq m \leq M$) are $\bm{v}_m = (x_m^p, 0, h)$, and the waveguide feed point is located at $\bm{v}_0 = (x_0^p, 0, h)$. By activating a subset of $N \leq M$ pinching points, the PWA focuses its radiation pattern toward the target receivers. We define the binary activation vector as:
\begin{equation}
    \bm{a} = [a_1, \dots, a_M]^T \in \{0,1\}^{M},
\end{equation}
where $a_m = 1$ if the $m$-th location is active, and $a_m = 0$ otherwise. The total power injected at the feed point is $P_0$. Assuming negligible propagation loss along the waveguide, the resulting per-antenna transmit power is $P_0/N$.

Due to the relatively short distances between the PWA and the receivers, the system operates in the radiative near-field. In this regime, the signal's amplitude and phase vary sharply over small spatial scales, potentially leading to performance fluctuations as receivers move. To ensure robust performance, we define service areas for both the EH and ID receivers. Each SA, denoted by $\mathcal{H}$, represents a spatial region centered at a target location $\bm{u}_{0,\mathcal{H}} = (x_{0,\mathcal{H}}, y_{0,\mathcal{H}}, 0)$ and includes all grid points $\bm{u} \in \mathcal{D}$ within a radius $\delta_{\mathcal{H}}$:
\begin{equation}
    \mathcal{H} = \left\{ \bm{u} \in \mathcal{D} \; \middle| \; \left\| \bm{u} - \bm{u}_{0,\mathcal{H}} \right\|_2 \le \delta_{\mathcal{H}} \right\}. 
\end{equation}

\subsection{Channel and Power Models}

Let $p_i$ denote the received power at a grid point $\bm{u}_i = (x_i, y_i, 0) \in \mathcal{H}$. The complex channel coefficient $h_{i,m}$ between the grid point $\bm{u}_i$ and the $m$-th candidate antenna location is determined by the Euclidean distance $d_{i,m} = \sqrt{(x_i - x_m^p)^2 + y_i^2 + h^2}$. Due to spherical wave propagation in near-field, the channel coefficient is modeled as: 
\begin{equation}
    h_{i,m} = \frac{\eta}{d_{i,m}} \, e^{-j \left( \frac{2\pi}{\lambda}d_{i,m} + \theta_m \right )},
\end{equation}
where $\eta = \lambda/(4\pi)$ is the free-space path loss constant, $\lambda$ is the free-space wavelength, and $\theta_m = \frac{2\pi}{\lambda_g} (x_m^p - x_0^p)$ is the phase shift due to waveguide propagation, $\lambda_g = \lambda/n_{\text{eff}}$ is the waveguide wavelength, and $n_{\text{eff}}$ the effective refractive index. 

The received power $p_i$ at grid point $\bm{u}_i \in \mathcal{H}$ is determined by superposition of signals radiated from active pinching antennas:
\begin{equation} \label{pi}
p_i \!=\! \frac{P_0}{N} \! \left| \sum_{m=1}^M \! h_{i,m} a_m \right|^2 \!\!=\! \frac{P_0}{N} \! \left| \bm{h}_{\mathcal{H},i}^H \bm{a} \right|^2 \!\!=\! \frac{P_0}{N} \bm{a}^T \bm{Q}_{\mathcal{H},i} \bm{a},
\end{equation}
where $\bm{h}_{\mathcal{H},i} = [h_{i,1}^*, \dots, h_{i,M}^*]^T$ is the channel coefficient vector specifically associated with grid point $\bm{u}_i$ within the SA $\mathcal{H}$, and $\bm{Q}_{\mathcal{H},i} = \bm{h}_{\mathcal{H},i} \bm{h}_{\mathcal{H},i}^H$ is the corresponding rank-one channel covariance matrix. The average received power $\bar{P}_{\mathcal{H}}$ is obtained by averaging $p_i$ across all $|\mathcal{H}|$ grid points within SA:
\begin{equation} \label{pn_avg}
\bar{P}_{\mathcal{H}} \!=\! \frac{1}{|\mathcal{H}|} \! \sum_{\bm{u}_i \in \mathcal{H}} \! p_i \!=\! \frac{P_0}{N |\mathcal{H}|} \! \left\| \bm{H}_{\mathcal{H}}^H \bm{a} \right\|^2 \!\!=\! \frac{P_0}{N |\mathcal{H}|} \bm{a}^T \bm{Q}_{\mathcal{H}} \bm{a},
\end{equation}
where $\bm{H}_{\mathcal{H}} = [\bm{h}_{\mathcal{H},1}, \bm{h}_{\mathcal{H},2}, \dots, \bm{h}_{\mathcal{H},|\mathcal{H}|}] \in \mathbb{C}^{M \times |\mathcal{H}|}$ is the channel matrix for the SA, whereas $\bm{Q}_{\mathcal{H}}$ is the associated spatial covariance matrix, defined by 
\begin{equation} \label{Qdef}
    \bm{Q}_{\mathcal{H}} = \bm{H}_{\mathcal{H}} \bm{H}_{\mathcal{H}}^H  \in \mathbb{C}^{M \times M}. 
\end{equation}
Note that $\bm{Q}_{\mathcal{H}}$ is a Hermitian positive semidefinite (PSD) matrix that characterizes the spatial correlation of channel responses over the SA $\mathcal{H}$. For small radii $\delta_{\mathcal{H}}$, the channel vectors within $\mathcal{H}$ are highly correlated, and $\bm{Q}_{\mathcal{H}}$ is approximately rank one. As $\delta_{\mathcal{H}}$ increases, the channel responses exhibit greater spatial diversity, thus increasing the rank of $\bm{Q}_{\mathcal{H}}$.


\section{PWA beamforming optimization}

In this section, we optimize the PWA activation vector to maximize the aggregate harvested power within the EH SA, subject to robust communication constraints for the ID receiver. 

\vspace{-3mm}

\subsection{Baseline: Absence of an Information Receiver} 
We first consider a scenario where the PWA transmitter serves an EH receiver in isolation. The goal is to maximize the total harvested power by the EH receiver deployed in the designated SA $\mathcal{H}_{\text{EH}}$ by optimally selecting the antenna activation vector $\bm{a}$:
\begin{equation} \label{MaxP}
\begin{aligned}
    (\text{OP1}) \quad \max_{\bm{a}} \quad & \bar P_\text{EH} = \bm{a}^T \bm{Q}_\text{EH} \bm{a} \\
    \text{s.t.} \quad & C1: \bm{a} \in \{0,1\}^{M}, \quad C2: \bm{1}^T \bm{a} = N,
\end{aligned}
\end{equation}
where $\bm{1} \in \mathbb{R}^{M}$ is the all-ones vector. Problem \eqref{MaxP} is a BQP problem subjected to a cardinality constraint $C2$, which enforces the activation of exactly $N$ pinching points. Due to the binary nature of $\bm{a}$, the problem is non-convex and NP-hard, rendering a global exhaustive search computationally intractable for large $M$. Next, we extend \eqref{MaxP} by imposing an achievable rate constraint for an ID receiver within its designated SA $\mathcal{H}_{\text{ID}}$, considering two design paradigms: ergodic and pointwise.

\vspace{-3mm}

\subsection{Ergodic Rate Constraint}
This formulation assumes the ID receiver moves within the SA $\mathcal{H}_{\text{ID}}$, causing the effective channel to vary with spatial position. In this view, the achievable rate corresponds to the ergodic capacity of a channel with random spatial states, where each grid point in $\mathcal{H}_{\text{ID}}$ represents one such state. The spatial average over all grid points therefore plays the role of the expectation in the ergodic capacity formula:
\begin{align} \label{avg_constraint}
    \bar R_{\text{ID}} = \frac{1}{|\mathcal{H}_{\text{ID}}|} \sum_{\bm{u}_j \in \mathcal{H}_{\text{ID}}} \log_2 \left(1 + \frac{p_j}{\sigma_N^2} \right), 
\end{align}
where $\sigma_N^2$ is the ID receiver's thermal noise power, and $p_j = (P_0/N) \bm{a}^T \bm{Q}_{\text{ID},j} \bm{a}$ (cf. \eqref{pi}). We thus add another constraint to (OP1) to ensure a fixed reliable communication rate: 
\begin{align} \label{OP2}
    (\text{OP2}) \quad \max_{\bm{a}} \quad & \bm{a}^T \bm{Q}_\text{EH} \bm{a} \notag \\
    \text{s.t.} \quad & C0: \bar R_{\text{ID}} \geq R_{th} \notag \\
    & C1: \bm{a} \in \{0,1\}^{M}, \quad C2: \bm{1}^T \bm{a} = N.
\end{align}
where $R_{th}$ is the target information rate at the ID receiver.

\vspace{-3mm}

\subsection{Pointwise Rate Constraint} 
To provide strict QoS guarantees, we require that the target rate $R_{th}$ be satisfied at each grid point within $\mathcal{H}_{\text{ID}}$. This provides a robust "fairness" guarantee, ensuring the receiver achieves rate of at least $R_{th}$ regardless of its specific location within the SA, which yields the following formulation: 

\begin{equation} \label{OP3}
\begin{aligned}
    (\text{OP3}) \quad \max_{\bm{a}} \quad & \bm{a}^T \bm{Q}_\text{EH} \bm{a} \\
    \text{s.t.} \quad & C0: \log_2 \left(1 + \frac{p_j}{\sigma_N^2} \right) \geq R_{th}, \, \forall \bm{u}_j \in \mathcal{H}_{\text{ID}} \\
    & C1: \bm{a} \in \{0,1\}^{M}, \quad C2: \bm{1}^T \bm{a} = N,
\end{aligned} 
\end{equation}
where $p_j = (P_0/N) \bm{a}^T \bm{Q}_{\text{ID},j} \bm{a}$ (cf. \eqref{pi}). Note that the OP3 constitutes a worst-case robust optimization over the spatial uncertainty set $\mathcal{H}_{\text{ID}}$, as it enforces $\min_{j \in \mathcal{H}_{\text{ID}}} R_j \geq R_{th}$. Together with the average-case formulation of OP2, these two problems establish the fundamental tradeoff between average-case and worst-case spatial robustness for the PWA systems. 

\section{Robust optimization framework}

\subsection{Semidefinite relaxation for upper bound characterization}

To overcome the intractability of OP1, OP2, and OP3, we employ semidefinite relaxation (SDR) to establish a rigorous upper bound on their optimal values. 

\subsubsection*{SDR for OP3} Let $\bm{X} = \bm{a}\bm{a}^T$ be the lifted matrix variable. We relax the binary condition $\bm{a} \in \{0,1\}^M$ to $\bm{X} \succeq 0$, $0 \le \bm{X}_{ii} \le 1$, and $\mathrm{Tr}(\bm{X})=N$. Let $J = |\mathcal{H}_{\rm ID}|$ denote the number of grid points in the ID SA. The quadratic rate constraint is linearized via the trace operation, 
\begin{align}
    \mathrm{Tr}(\bm{Q}_{{\rm ID},j} \bm{X}) \ge \gamma_{th}, 
\end{align} 
for each point $\bm{u}_j \in \mathcal{H}_{\mathrm{ID}}$, where $\gamma_{th} = \sigma_N^2 (2^{R_{\rm th}} - 1)$ is the required power threshold. The SDR of OP3 is formulated as:
\begin{align} \label{eq:OP3_SDR_refined}
\max_{\bm{X} \succeq 0} \quad
    & F^* = \mathrm{Tr}(\bm{Q}_{\mathrm{EH}} \bm{X}) \\
\text{s.t.} \quad
    & \mathrm{Tr}(\bm{Q}_{{\rm ID},j} \bm{X}) \ge \gamma_{th}, 
      \quad j=1,\ldots,J, \nonumber \\
    & \mathrm{Tr}(\bm{X})=N, \quad 0 \le \bm{X}_{ii} \le 1,\;\; \forall i. \nonumber
\end{align}

\subsubsection*{SDR for OP2 (including Jensen's Inequality)} For OP2, the constraint is specified on the spatial average rate. Direct relaxation is difficult due to the summation of logarithms. To obtain a tractable convex upper bound, we invoke Jensen’s inequality, $\frac{1}{J}\sum \log(1+z_j) \le \log(1+\frac{1}{J}\sum z_j)$. By replacing the average rate with the log of the average SNR, we expand the feasible set and obtain the linear surrogate constraint:
\begin{align}
    \mathrm{Tr}(\bar{\bm{Q}}_{\rm ID} \bm{X}) \ge \gamma_{th}, 
\end{align}
where $\bar{\bm{Q}}_{\rm ID} = (1/J) \sum_{j=1}^J \bm{Q}_{{\rm ID},j} = (1/J) \bm{Q}_{\rm ID}$ is the averaged spatial covariance matrix of SA $\mathcal{H}_{\text{ID}}$ (c.f. (\ref{Qdef})). This collapses the $J$ individual point constraints of (\ref{eq:OP3_SDR_refined}) into a single constraint.  

\subsubsection*{SDR for OP1} Removing all rate constraints yields:
\begin{align}
    \max_{\bm{X} \succeq 0} \; \mathrm{Tr}(\bm{Q}_{\rm EH} \bm{X}) \quad\text{s.t.}\quad \mathrm{Tr}(\bm{X})=N,\;\; 0 \le \bm{X}_{ii} \leq 1.
\end{align}

\subsubsection{Dual Problem Formulation} To determine the SDR optimal value $F^\star$, we derive the dual problem of the primal problem~\eqref{eq:OP3_SDR_refined}, which avoids solving the full SDP. Let us introduce the Lagrangian dual variables: $\mu \in \mathbb{R}$ for the trace equality $\mathrm{Tr}(\bm{X})=N$, $\bm{d} = [d_1, \dots, d_M]^T \ge \bm{0}$ for the diagonal constraints $\bm{X}_{ii} \le 1$, and $\bm{\nu} = [\nu_1, \dots, \nu_J]^T \ge \bm{0}$ for the pointwise rate constraints. Let $\bm{D} = \mathrm{diag}(\bm{d})$. The Lagrangian of \eqref{eq:OP3_SDR_refined} is given by:

\begin{equation}
    \mathcal{L}(\bm{X}, \mu, \bm{\nu}, \bm{d}) = \mathrm{Tr} \left( \bm{A}(\bm{\nu}, \bm{d}, \mu) \bm{X} \right) + N \mu + \mathbf{1}^\top \bm{d} - \gamma_{th} \sum_{j=1}^J \nu_j,
\end{equation}
where the aggregate matrix $\bm{A}(\bm{\nu}, \bm{d}, \mu)$ is defined by 
\begin{equation} \label{psd_neg}
    \bm{A}(\bm{\nu}, \bm{d}, \mu) = \bm{Q}_{\rm EH} + \sum_{j=1}^J \nu_j \bm{Q}_{{\rm ID},j} - \bm{D} - \mu \bm{I}. 
\end{equation}
The Lagrangian dual function is obtained by maximizing over $\bm{X} \succeq 0$:
\begin{equation} \label{dualfunc}
    g(\mu, \bm{\nu}, \bm{d}) = \sup_{\bm{X} \succeq 0} \mathcal{L}(\bm{X}, \mu, \bm{\nu}, \bm{d}).
\end{equation}
Since the Lagrangian is affine in $\bm{X}$, the supremum is finite iff the coefficient matrix is negative semidefinite, i.e., $\bm{A}(\bm{\nu}, \bm{d}, \mu) \preceq 0$. If this condition holds, the optimal $\bm{X}$ is $\bm{0}$, yielding $\sup_{\bm{X}} \mathrm{Tr}(\bm{A}\bm{X}) = 0$. Otherwise, the dual function is unbounded ($+\infty$). Condition $\bm{A}(\bm{\nu}, \bm{d}, \mu) \preceq 0$ is equivalent to:
\begin{equation} \label{matrix_inequality}
    \bm{Q}_{\rm EH} + \sum_{j=1}^J \nu_j \bm{Q}_{{\rm ID},j} - \bm{D} \preceq \mu \bm{I}.
\end{equation}
To obtain the tightest dual bound, we select the smallest feasible $\mu$, which corresponds to the maximum eigenvalue of the matrix on the left-hand side of (\ref{matrix_inequality}), 
\begin{equation}
    \mu^\star = \lambda_{\max} \Big( \bm{Q}_{\rm EH} + \sum_{j=1}^J \nu_j \bm{Q}_{{\rm ID},j} - \bm{D} \Big).
\end{equation}
Substituting $\mu^\star$ back into the dual objective, and invoking strong duality, the optimal primal value $F^\star$ is obtained by minimizing the dual function over the remaining multipliers:
\begin{align} \label{ub_diag}
    F^\star = \min_{\bm{\nu} \ge 0,\, \bm{d} \ge 0} \Big( 
    N \mu^\star + \mathbf{1}^\top \bm{d} - \gamma_{th} \sum_{j=1}^J \nu_j \Big).
\end{align}
Eq.~\eqref{ub_diag} specializes to OP2 by replacing summation with a single term $\nu_0 Q_{\rm ID}$, and OP1 by setting $\bm{\nu} = \bm{0}$.

\subsubsection{Nesterov-smoothed dual solver}

The dual objective in \eqref{ub_diag} involves the spectral term $\lambda_{\max}(\cdot)$, which is non-smooth and precludes standard gradient methods. We address this by employing Nesterov smoothing approximation of $\lambda_{\max}(\bm{B})$ as:
\begin{align} \label{eq:smooth_lmax_def} 
    f_\tau(\bm{B}) = \tau \log \! \left( \mathrm{Tr}\!\left(\exp(\bm{B}/\tau)\right) \right), 
\end{align}
where $\tau > 0$ is the smoothing parameter. This approximation satisfies: $\lambda_{\max}(\bm{B}) \le f_\tau(\bm{B}) \le \lambda_{\max}(\bm{B}) + \tau \log M$. By setting $\tau = \epsilon / \log M$, we ensure approximation error is controlled within a desired tolerance $\epsilon$. The smoothed objective function is convex and differentiable with a Lipschitz constant $L = N/\tau$. The gradient takes the form of a normalized matrix exponential:
\begin{align} \label{eq:smooth_grad}
    \bm{G} = \nabla_{\bm{B}} f_\tau(\bm{B}) = \frac{\exp(\bm{B}/\tau)}{\mathrm{Tr}\!\left(\exp(\bm{B}/\tau)\right)},
\end{align}
where $\bm{G} \succeq 0$ and $\mathrm{Tr}(\bm{G})=1$. This structure admits the use of \textit{projected gradient descent} (PGD), as detailed in Algorithm~\ref{alg:pgd_dual_OP3}. The partial derivatives of the smoothed dual function $g_{\tau}(\bm{\nu},\bm{d}) \triangleq N f_\tau\!\big(\bm{Q}_{\rm EH} + \sum_{j} \nu_j \bm{Q}_{{\rm ID},j} - \mathrm{diag}(\bm{d})\big) + \mathbf{1}^\top \bm{d} - \gamma_{th} \sum_{j} \nu_j$ with respect to $\bm{\nu}$ and $\bm{d}$ are given by:
\vspace{-7mm}
\begin{align}
    \delta_{\nu,j} \triangleq \frac{\partial g_\tau}{\partial \nu_j}
    &= N \mathrm{Tr}(\bm{G} \bm{Q}_{\mathrm{ID},j}) - \gamma_{th}, \quad j = 1,\ldots,J,\\
    \delta_{d,i} \triangleq \frac{\partial g_\tau}{\partial d_i}
    &= 1 - N [\bm{G}]_{ii}, \quad i = 1,\ldots,M. 
\end{align}

\begin{algorithm}[t]
\caption{Gradient Descent for Solving \eqref{ub_diag}}
\label{alg:pgd_dual_OP3}
\begin{algorithmic}[1]
\REQUIRE $\bm{Q}_{\mathrm{EH}}$, $\{\bm{Q}_{{\mathrm{ID}},j}\}_{j=1}^{J}$, $N$, $\gamma_{th}$, $\epsilon$, max iters $T$.
\STATE \textbf{Initialize:} $\bm{\nu}^{(0)} \in \mathbb{R}_{+}^{J}$, $\bm{d}^{(0)} \in \mathbb{R}_{+}^{M}$, $F^\star \leftarrow +\infty$.
\STATE \textbf{Set:} $\tau \leftarrow \epsilon/\log(M)$, step size $\eta \leftarrow \tau/N$.
\FOR{$t=0,1,\dots,T-1$}
    \STATE $\bm{B}^{(t)} \leftarrow \bm{Q}_{\mathrm{EH}} + \sum_{j=1}^{J} \nu_{j}^{(t)} \bm{Q}_{{\mathrm{ID}},j} - \mathrm{diag}(\bm{d}^{(t)})$
    \STATE $\bm{Z}^{(t)} \leftarrow \exp(\bm{B}^{(t)}/\tau)$, \quad $\bm{G}^{(t)} \leftarrow \bm{Z}^{(t)} / \mathrm{Tr}(\bm{Z}^{(t)})$ 
    \STATE $F^{(t)} \leftarrow N \tau \log(\mathrm{Tr}(\bm{Z}^{(t)})) + \mathbf{1}^\top \bm{d}^{(t)} - \gamma_{th}\sum_{j=1}^{J} \nu_{j}^{(t)}$
    \IF{$F^{(t)} < F^\star$}
        \STATE $F^\star \leftarrow F^{(t)}$, \quad $(\bm{\nu}^\star, \bm{d}^\star) \leftarrow (\bm{\nu}^{(t)}, \bm{d}^{(t)})$
    \ENDIF
    \STATE \textbf{Compute Gradients:}
    \STATE \quad $\delta_{\nu,j}^{(t)} \leftarrow N \mathrm{Tr}(\bm{G}^{(t)}\bm{Q}_{{\mathrm{ID}},j}) - \gamma_{th}, \quad \forall j$
    \STATE \quad $\delta_{d,i}^{(t)} \leftarrow 1 - N [\bm{G}^{(t)}]_{ii}, \quad \forall i$
    \STATE \textbf{Projected Update:}
    \STATE \quad $\bm{\nu}^{(t+1)} \leftarrow \max \big\{\mathbf{0},\, \bm{\nu}^{(t)} - \eta \bm{\delta}_{\bm{\nu}}^{(t)}\big\}$
    \STATE \quad $\bm{d}^{(t+1)} \leftarrow \max \big\{\mathbf{0},\, \bm{d}^{(t)} - \eta \bm{\delta}_{\bm{d}}^{(t)}\big\}$
\ENDFOR
\RETURN $F^\star$ and $(\bm{\nu}^\star, \bm{d}^\star)$
\end{algorithmic}
\end{algorithm}

\subsubsection{Low-rank approximation and complexity analysis}
The dominant computational cost in Algorithm~\ref{alg:pgd_dual_OP3} is the evaluation of the matrix exponential $\exp(\bm{B}/\tau)$, which generally requires $\mathcal{O}(M^3)$ operations via eigendecomposition. While this scales cubically, it is strictly superior to solving the primal SDR directly. Primal interior-point methods scale as $\mathcal{O}(n^{3.5})$ where $n$ is the number of variables. Since the primal variable $\bm{X}$ has $n = M^2$ entries, the complexity scales as $\mathcal{O}(M^7)$, which is prohibitive for large $M$.

To further reduce the cost, we employ a Lanczos-based low-rank approximation. This reduces the evaluation of $\exp(\bm{B}/\tau)$ to operations on a small tridiagonal matrix $\bm{T}_r \in \mathbb{R}^{r \times r}$, where $r \ll M$. The smoothed gradient is then approximated as $\bm{G} \approx \hat{\bm{V}}_r \exp(\bm{\Lambda}_r) \hat{\bm{V}}_r^\top / \mathrm{Tr}(\exp(\bm{\Lambda}_r))$, where $\hat{\bm{V}}_r$ represents the approximate eigenvectors generated by $r$ matrix-vector products. This refinement reduces the per-iteration spectral complexity from $\mathcal{O}(M^3)$ to $\mathcal{O}(rM^2)$. For OP1 and OP2, this spectral step dominates the runtime. 

For OP3, a naive gradient update would cost $\mathcal{O}(JM^2)$, potentially exceeding the spectral cost if the grid size $J$ is large (i.e., $J > r$). To prevent this bottleneck, we exploit the low-rank structure of $\bm{G}$ to compute the trace terms $\mathrm{Tr}(\bm{G} \bm{Q}_{\mathrm{ID},j}) = \bm{h}_{\mathrm{ID},j}^H \bm{G} \bm{h}_{\mathrm{ID},j}$ via efficient matrix-vector products, which requires only $\mathcal{O}(rM)$ operations per grid point. The total complexity for OP3 thus becomes $\mathcal{O}(rM^2 + JrM)$. 

\subsection{Discrete Antenna Selection via Swap-Based Local Search} 

While the SDR framework provides rigorous upper-bound benchmarks, it does not directly produce a realizable binary activation vector $a$. To address this limitation, and motivated by [\ref{swipt_1_lit}], we adopt a computationally efficient swap-based local search algorithm to obtain feasible binary activation patterns. The algorithm starts from an initial feasible subset of $N$ antennas and iteratively refines the selection to maximize the harvested energy. At each iteration, the local neighborhood is defined by all valid single-element swaps, where one active antenna is deactivated and one inactive candidate is activated. A swap is accepted only if it yields a strict improvement in the harvested energy while maintaining feasibility with the rate constraints. This greedy improvement strategy guarantees monotonic convergence to a locally optimal binary solution.

The computational efficiency of this search relies on the linear property of the trace operator. Since both the harvested and received powers at a grid point can be expressed as sums of individual antenna contributions, the effect of a candidate swap can be evaluated via incremental updates. As a result, each swap evaluation requires constant complexity per grid point, leading to a per-pass computational complexity of $\mathcal{O}(N(M-N)J)$. In practice, convergence is achieved within a small number of passes due to the monotonic improvement property and the favorable spectral structure of near-field PWA channels.

\vspace{2mm}

\section{Numerical Results}

In this section, we evaluate the performance of the proposed robust antenna selection schemes for a near-field SWIPT system. The PWA operates at $f_c = 28$~GHz and comprises a dielectric waveguide of length $6$~m with an effective refractive index $n_{\mathrm{eff}} = 1.4$, deployed at a height $h = 3$~m. The inter-element spacing is fixed to $\Delta = \lambda$, yielding $M = 560$ candidate pinching locations. The inter-element spacing is fixed to $\Delta = \lambda$, yielding $M = 560$ candidate pinching locations. From these, a subset of $N = 40$ antennas is activated to cover a $6 \times 6$~m square region $\mathcal{D}$. The total transmit power is $P_0 = 10$~W, and the receiver noise power is $\sigma_N^2 = 10^{-12}$~W. 

We consider a \emph{time-sharing point-based benchmark} scheme representative of the standard PWA design approach~[\ref{flexible_lit}]--[\ref{swipt_2_lit}], where antenna selection is independently optimized for each receiver at its nominal SA center. The EH and ID transmissions are separated in time. The harvested power and achievable rate are averaged across the SAs to reflect spatial uncertainty. 

\subsection{Rate-Energy Trade-off and Duality Gap}

\begin{figure}[t]
    \centering
    \includegraphics[width=0.47\textwidth]{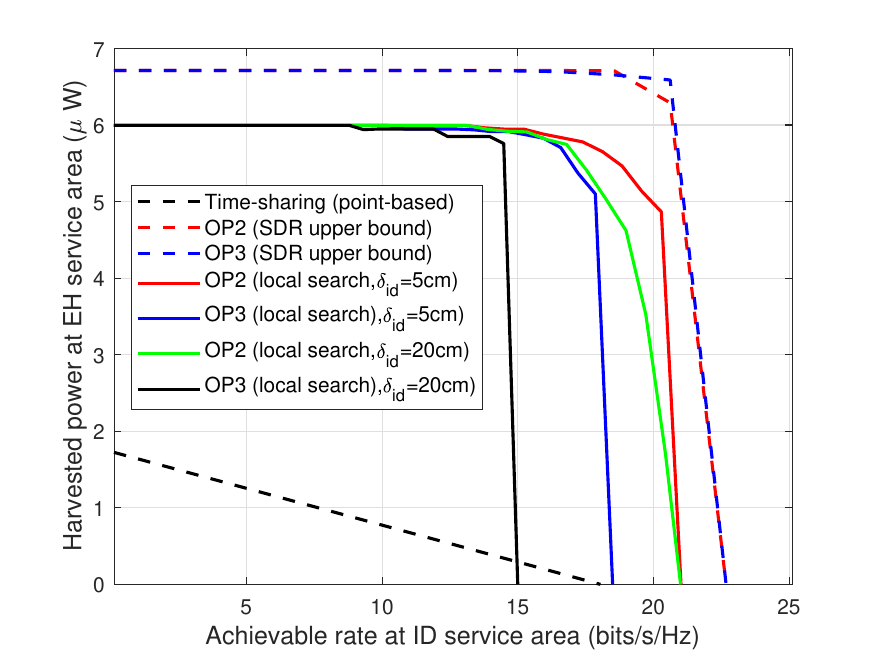} 
    \caption{Rate-energy region boundaries of the PWA system.} 
    \label{fig:RE_region}
    \vspace{-3mm}
\end{figure}

Fig.~\ref{fig:RE_region} illustrates the rate-energy (R-E) region boundaries for the ergodic (OP2) and pointwise (OP3) formulations. The dashed lines represent the theoretical upper bounds derived via SDR for the large SA scenario ($\delta_{\rm EH} = \delta_{\rm ID} = 20$~cm). The solid curves represent the realizable integer solutions obtained via swap-based local search for two ID SA radii. The red and blue curves apply to a small radius of $\delta_{\rm ID} = 5$~cm, while the green and black curves apply to a larger radius of $\delta_{\rm ID} = 20$~cm. The time-sharing benchmark (dashed black line) lies far below all proposed R-E curves, achieving only $\bar{P}_{\text{EH}} \approx 1.74\,\mu$W at zero rate. The gap reflects the fragility of point-based near-field beamforming: while the focused beam performs well at target coordinate, it degrades rapidly when averaged across SA. 

\begin{figure}[t]
    \centering
    \includegraphics[width=0.47\textwidth]{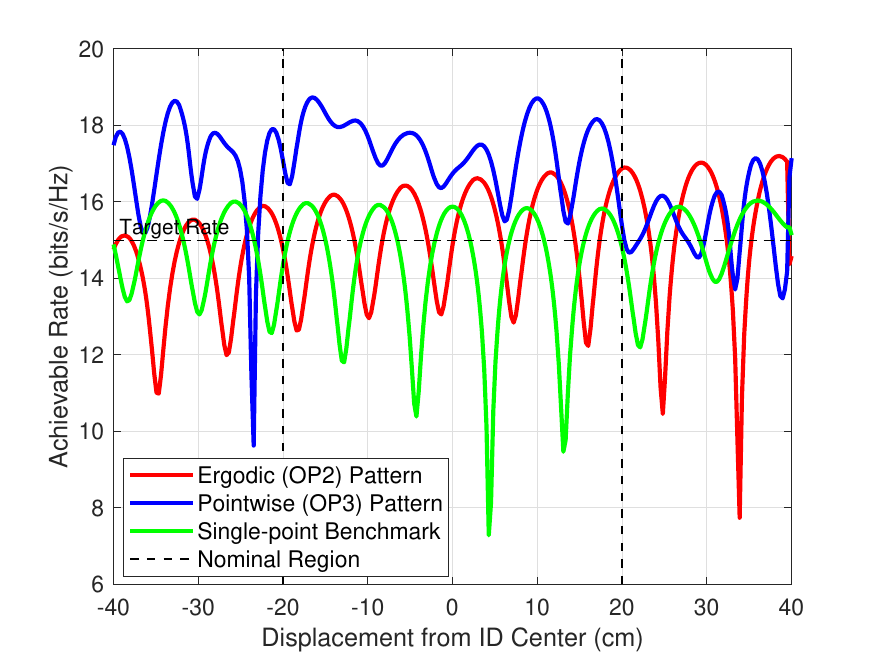} 
    \caption{Spatial resilience of achievable rate against ID receiver's displacement.}
    \label{fig:spatial_resilience}
    \vspace{-3mm}
\end{figure}

Actually, OP2 treats the user position as a random variable for attaining the average ID rate within the predefined SA, whereas OP3 treats it as an unknown parameter within the SA. Consequently, a significant performance trade-off is evident between the two designs. For small radii ($\delta_{\rm ID}=5$~cm), the gap between the ergodic design (red) and the robust design (blue) is only about $2$~bits/s/Hz. However, for large radii ($\delta_{\rm ID}=20$~cm), the gap between the ergodic design (green) and the robust design (black) reaches about $5$~bits/s/Hz. This quantifies the ``cost of robustness,'' showing that as the SA expands, enforcing strict QoS under OP3 substantially decreases the achievable rate compared to the OP2 design. 

Comparing the $20$~cm case (green and black curves) against their corresponding SDR upper bounds (dashed lines), we observe a relative energy gap of roughly $12\%$ in the saturation regime ($6.8~\mu$W versus $6.0~\mu$W). This gap stems from the penalty of discretization, as the SDR permits fractional antenna activation patterns that cannot be implemented with PWAs.

\subsection{Spatial Resilience and Outage Analysis}

Fig.~\ref{fig:spatial_resilience} evaluates the spatial resilience of the ID receiver as it moves linearly through the SA center ($\pm 40$~cm). Activation vectors $\bm{a}^\star$ are selected corresponding to the ``knee'' of the $R$--$E$ region for a target rate $R_{th} = 15$~bits/s/Hz (c.f. black curve in Fig.~\ref{fig:RE_region}). Consistent with the R-E region results, the point-based benchmark in Fig.~\ref{fig:spatial_resilience} confirms that point-based designs exhibit sensitivity to displacement, with a minimum rate of only $7.28$~bits/s/Hz within the SA versus $15.44$~bits/s/Hz for the OP3 design. 

The ergodic design (OP2, red curve) achieves a spatial average rate of $15.14$~bits/s/Hz, satisfying $\bar R_{\text{ID}} \ge R_{th}$. However, the rate exhibits high variance with displacement. Fluctuations occur even within the SA, where the local rate occasionally dips below the target. These fades confirm that the ergodic formulation cannot guarantee consistent service quality under spatial uncertainty. In contrast, the pointwise design (OP3, blue curve) synthesizes a much more uniform rate profile by spreading signal energy across the SA. It maintains a minimum rate of $15.44$~bits/s/Hz throughout the entire SA, strictly satisfying the QoS requirement $\min_{j\in \mathcal{H}_{\text{ID}}} R_j \ge R_{th}$. While some fluctuation remains, the spatial stability is significantly improved; a deep fade is observed only at $\approx -23$~cm, which is outside the SA. This confirms that the pointwise design successfully eliminates coverage holes within the desired SA.

\section{Conclusion}

This paper proposed a robust design framework for near-field SWIPT using PWAs. To address user mobility and mitigate the rapid signal fluctuations inherent to the radiative near field, we defined service areas around each receiver where performance is guaranteed through either spatial averaging or strict pointwise constraints. The fundamental energy-rate tradeoff using SDR benchmarks revealed that optimal solutions exploit the strong spectral concentration of near-field channels. Comparisons with a point-based benchmark confirm that the observed performance gains are attributable to the SA-based formulation rather than the specific optimization algorithm, underscoring that spatially robust design is essential for realizing the full potential of near-field PWA systems.

\bibliographystyle{IEEEtran}

\end{document}